\documentclass[a4paper,11pt]{article}
\usepackage{pos}
\pdfoutput=1

\title{\boldmath $\mathcal{O}(\alpha_s^3)$ calculations for the inclusive determination of $|V_{cb}|$}
\ShortTitle{$\mathcal{O}(\alpha_s^3)$ calculations for the inclusive determination of $|V_{cb}|$}

\author*[a]{Kay Schönwald}

\affiliation[a]{Institut für Theoretische Teilchenphysik, Karlsruhe Institute of Technology (KIT), 76128 Karlsruhe, Germany}

\emailAdd{kay.schoenwald@kit.edu}

\abstract{
  For the determination of the Cabbibo-Kobayashi-Maskawa matrix element $|V_{cb}|$ from 
  inclusive data a precise knowledge of the semileptonic $b \to c$ decay rate is necessary.
  Since this observable has a bad convergence behavior when the heavy quark 
  masses are expressed in the on-shell or $\overline{\text{MS}}$ scheme the latest determinations 
  have been obtained in the so called kinetic mass scheme.
  The relation between the different schemes needs to be known to high precision as well.
  In this proceedings we present our recent calculations which push the precision of both 
  ingredients to $\mathcal{O}(\alpha_s^3)$. 
  The results can be used to improve the inclusive determination of $|V_{cb}|$.     
}

\FullConference{%
  11th International Workshop on the CKM Unitarity Triangle (CKM2021)\\
  22-26 November 2021\\
  The University of Melbourne, Australia
}

\begin{document}
\maketitle

\section{Introduction}

Inclusively the Cabbibo-Kobayashi-Maskawa (CKM) matrix elements $|V_{ub}|$ and $|V_{cb}|$ are 
extracted from global fits to experimental data on the semileptonic 
$B \to X_{c(u)} \ell \overline{\nu}$ decay width and moments of several kinematic distributions 
like the ones for the hadronic invariant mass or the lepton energy 
\cite{Bauer:2004ve,Gambino:2013rza,Alberti:2014yda,Gambino:2016jkc,Bordone:2021oof}.
Theoretically these decays can be described in the heavy quark effective theory (HQET) as 
a double expansion in the strong coupling constant $\alpha_s$ and the inverse heavy (bottom)
quark mass $1/m_b$.
The leading term in the expansion in the heavy quark mass is given by the free quark decay
$b \to c(u) \ell \overline{\nu}$.
The convergence of this double series depends crucially on the scheme used to express the 
heavy quark mass.
Here, the pole mass suffers from renormalon ambiguities \cite{Beneke:1994sw,Bigi:1994em}, 
which can be avoided by going to, for example, the $\overline{\text{MS}}$ mass scheme, 
which is often employed in LHC analyses.
However, at low energies it is advantageous to switch to so called threshold masses like the 
$1S$ \cite{Hoang:1998ng,Hoang:1998hm,Hoang:1999us} or the kinetic \cite{Bigi:1994ga,Bigi:1996si} mass scheme.  

In these proceedings we summarize the recent calculation of the $\mathcal{O}(\alpha_s^3)$ 
relation between the pole and kinetic heavy quark mass (c.f.~Refs.~\cite{Fael:2020iea,Fael:2020njb}) and of the 
semileptonic decay rate (c.f.~Ref.~\cite{Fael:2020tow}) as well as their phenomenological implications.

\section{\boldmath The Kinetic Heavy Quark Mass to $\mathcal{O}(\alpha_s^3)$}

The kinetic heavy quark mass is defined in strong analogy to the relation between the mass 
of a heavy meson $M_H$ and the respective heavy quark mass $m_Q$:
\begin{align}
  M_H &= m_Q + \overline{\Lambda} + \frac{\mu_\pi^2}{2 m_Q} + \mathcal{O}\left( \frac{1}{m_Q^2} \right) ~,
  \label{eq:1}
\end{align}
where the parameter $\mu_\pi^2$ is a non-perturbative matrix elements of local 
HQET operators and $\overline{\Lambda}$ is the binding energy of the meson in the heavy quark 
limit.
The relation between the kinetic heavy quark mass and the pole (or equivalently on-shell) mass  
is obtained from Eq.~\eqref{eq:1} by identifying $M_H \to m_Q^{\rm OS}$, $m_Q \to m_Q^{\rm kin}$
and evaluating the operator matrix elements in perturbation theory \cite{Bigi:1996si}.
The explicit relation up to $\mathcal{O}(1/m_Q)$ reads:
\begin{align}
  m_Q^{\rm OS} &= m_Q^{\rm kin}(\mu) 
    + [\overline{\Lambda}]_{\rm pert}
    + \frac{[\mu_\pi^2(\mu)]_{\rm pert}}{2 m_Q^{\rm kin}(\mu)}
    + \mathcal{O}\left( \frac{1}{m_Q^2} \right)
    ~.
    \label{eq:2}
\end{align}
The relation has been previously calculated up to $\mathcal{O}(\alpha_s^2)$ 
\cite{Bigi:1996si,Czarnecki:1997sz}.

\begin{figure}
  \centering 
  \includegraphics[width=0.32\textwidth]{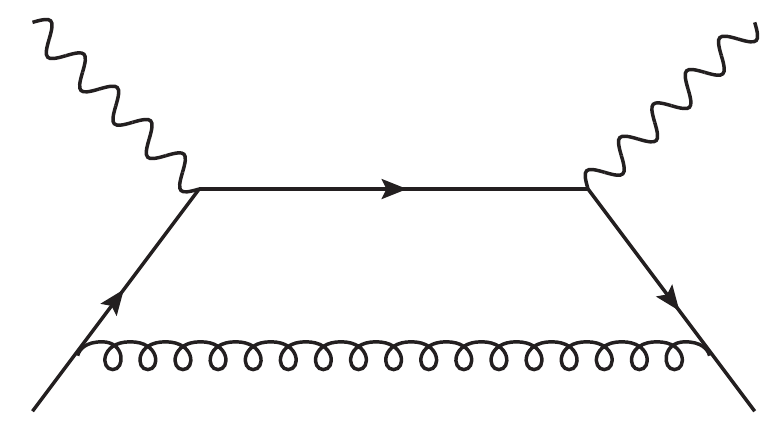}
  \includegraphics[width=0.32\textwidth]{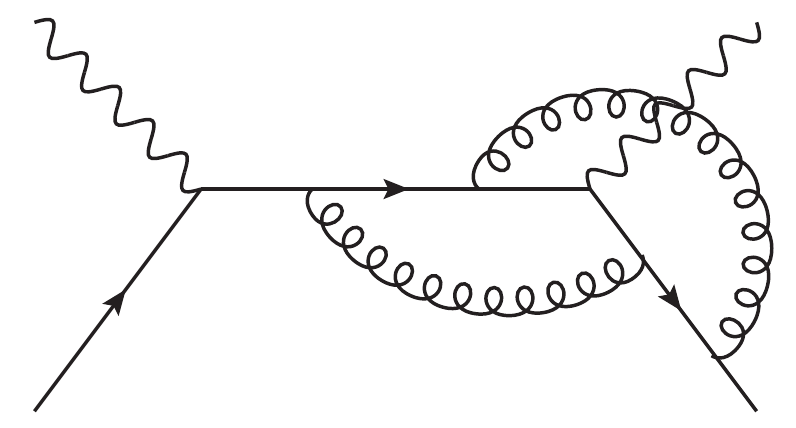}
  \includegraphics[width=0.32\textwidth]{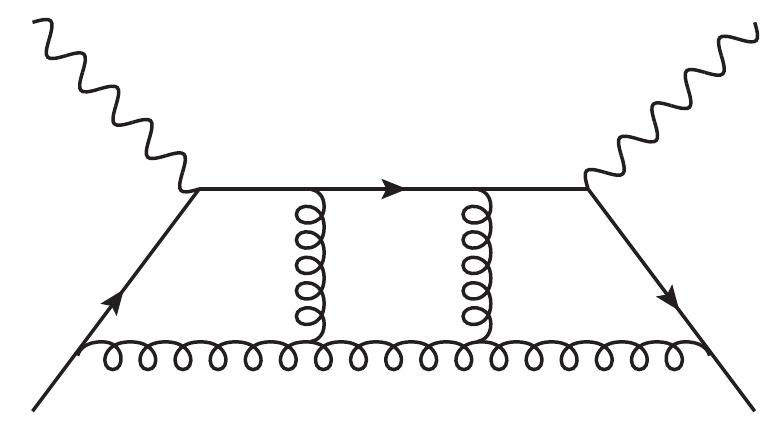}
  \caption{Sample Feynman diagrams for the scattering of an external current
  (wavy line) on a heavy quark (solid line). Taken from Ref.~\cite{Fael:2020njb}.}
  \label{fig:1}
\end{figure}

A constructive way to compute the HQET parameters in perturbation is given by the Small Velocity (SM)
sum rules \cite{Bigi:1994ga}.
Here, one considers the scattering of a heavy quark $Q$ on a current $J$.
The current transfers energy to the quark $Q$ and excites it, causing possibly
emissions of further gluons or quarks. 
We denote the inclusive final state as $X_Q$.
Working in the rest frame of the initial heavy quark we can define the 
excitation energy $\omega$ by 
\begin{align}
  \omega &= q_0 - q_0^{\rm min} = q_0 - \left( \sqrt{\vec{q}^2 + m_Q^2} - m_Q \right) ~,
\end{align}
where $q=(q_0,\vec{q})$ is the 4-momentum of the current.
The velocity of the system $X_Q$ after the scattering is given by $\vec{v}=\vec{q}/m_Q$.
The perturbative versions of the operator matrix elements can then be given by 
\begin{align}
  [\overline{\Lambda}]_{\rm pert} &= 
    \lim\limits_{\vec{v} \to 0}
    \lim\limits_{m_Q \to \infty}
    \frac{2}{\vec{v}^2} \frac{\int_0^\mu \omega \, W(\omega,\vec{v}) d\omega}{\int_0^\mu W(\omega,\vec{v}) d\omega}
  ~, 
  \label{eq:3} \\
  [\mu_\pi^2]_{\rm pert} &=
    \lim\limits_{\vec{v} \to 0}
    \lim\limits_{m_Q \to \infty}
    \frac{3}{\vec{v}^2} \frac{\int_0^\mu \omega^2 \, W(\omega,\vec{v}) d\omega}{\int_0^\mu W(\omega,\vec{v}) d\omega}
  ~,
  \label{eq:4}
\end{align}
where $W(\omega,\vec{v})$ is the structure function corresponding to the scattering and
the parameter $\mu$ is introduced as a Wilsonian cut-off in order to separate low and high energy effects.
The perturbative versions of the operator matrix elements are therefore given by moments 
of the scattering cross section.
However, the non-relativistic description in terms of excitation energy and velocity given 
in Eqs.~\eqref{eq:3} and \eqref{eq:4} do not allow a straight forward expansion on the level 
of Feynman diagrams.
For the calculation we followed the following strategy (see Ref.~\cite{Fael:2020njb} for a more detailed 
discussion):
\begin{itemize}
  \item We utilize the optical theorem and consider the discontinuity of the $bJ \to bJ$ 
  forward scattering diagrams (see Figure~\ref{fig:1} for example diagrams).
  \item We express the non-relativistic quantities $\omega$ and $\vec{v}^2$ in terms 
  of the Lorentz invariants 
  \begin{align}
    y = m_Q - s = - m_Q \omega (2+\vec{v}^2) + \mathcal{O}(\omega^2,\vec{v}^4)
    ~, \label{eq:5} \\
    q^2 = - m_Q \vec{v}^2 (m_Q - \omega) + \mathcal{O}(\omega^2,\vec{v}^4) 
    ~. \label{eq:6}
  \end{align}
  \item The limit $m_Q \to \infty$ can now be realized as the asymptotic expansion around 
  the threshold $s = m_Q^2$ (or equivalently $y = 0$) for which we use the strategy of 
  expansion by region \cite{Beneke:1997zp,Smirnov:2012gma}.
  The limit $\vec{v} \to 0$ can subsequently be realized by a naive Taylor expansion in $q$.
  When the leading terms in $y$ and $q^2$ of the structure function have been extracted 
  we use Eqs.~\eqref{eq:5} and \eqref{eq:6} to go back to the non-relativistic quantities and re-expand. 
\end{itemize}
This strategy now allows to use the full machinery of multi-loop calculations, i.e. 
we generate one-, two- and three-loop forward scattering diagrams with 
\texttt{qgraf} \cite{Nogueira:1991ex} and use \texttt{FORM} \cite{Ruijl:2017dtg} to insert the Feynman rules, perform 
the Dirac and color algebra and expand all loop-momenta according to the rules 
of asymptotic expansion. 
In the present case the momenta can either scale hard ($k_i \sim m_b$) or 
ultrasoft ($k_i \sim y/m_b$).
The corresponding regions have been cross-checked with the program \texttt{Asy.m} \cite{Jantzen:2012mw}.
After the expansion the denominators become linearly dependent, so a partial fraction 
decomposition becomes necessary. 
For this we used the program \texttt{LIMIT} \cite{Herren:2020ccq}, which automatizes this step.
This program also maps each scalar integral to a unique integral family, so that 
we were able to reduce all integrals to a small set of master integrals 
using the programs \texttt{FIRE} \cite{Smirnov:2019qkx} and \texttt{LiteRed} \cite{Lee:2013mka}.

If the loop momenta in the asymptotic expansion scale hard ($k_i \sim m_b$), the 
master integrals are given by on-shell propagator integrals, which are well studied 
in the literature \cite{Laporta:1996mq,Melnikov:2000zc,Lee:2010ik}.
For ultrasoft momenta ($k_i \sim y/m_b$) new types of master integrals appear which 
were evaluated using Mellin-Barnes 
%--- short
techniques and differential equations in auxillary parameters.
%--- long ---
% and integrals together with analytic summation techniques 
%implemented in the package \texttt{Sigma} \cite{}, differential equations in auxillary parameters,
%where we used the packages \texttt{OreSys} \cite{} and \texttt{HarmonicSums} \cite{} and 
%high precision numerical evaluations together with \texttt{PSLQ} \cite{pslq}.  
%For the analytic continuation in the $\epsilon = (4-d)/2$ we made use of the 
%\texttt{MB} package \cite{Czakon:2005rk,Smirnov:2009up}.

The final result is given by 
\begin{align}
  \frac{m^{\text{kin}}}{m^{\text{OS}}} &=
  1
  - \frac{\alpha_s^{(n_l)}}{\pi} C_F
  \biggl(
    \frac{4}{3} \frac{\mu}{m^{\text{OS}}} 
                     + \frac{1}{2} \frac{\mu^2}{\left({m^{\text{OS}}}\right)^2}
  \biggr)
  + \left( \frac{\alpha_s^{(n_l)}}{\pi}\right)^2 C_F
  \Biggl\{
    \frac{\mu}{{m^{\text{OS}}}}
    \Biggl[
      C_A 
      \biggl(
        - \frac{215}{27}
        + \frac{2\pi^2}{9} 
        + \frac{22}{9} l_\mu
      \biggr)
      \nonumber \\ & \mbox{}
      + n_l T_F 
      \biggl(
        \frac{64}{27}
        - \frac{8}{9} l_\mu
      \biggr)
    \Biggr]
    + \frac{\mu^2}{\left({m^{\text{OS}}}\right)^2}
    \Biggl[
      C_A 
      \biggl(
        - \frac{91}{36}
        + \frac{\pi^2}{12}
        + \frac{11}{12} l_\mu
      \biggr)
      + n_l T_F
      \biggl(
        \frac{13}{18}
        - \frac{1}{3} l_\mu
      \biggr)
    \Biggr]
  \Biggr\}
  \nonumber \\ & \mbox{}
  + \left( \frac{\alpha_s^{(n_l)}}{\pi} \right)^3 C_F
  \Biggl\{
    \frac{\mu}{{m^{\text{OS}}}}
    \Biggl[
      C_A^2
      \biggl(
        - \frac{130867}{1944}
        + \frac{511 \pi^2}{162}
        + \frac{19 \zeta_3}{2} 
        - \frac{\pi^4}{18}
        + \biggl(
          \frac{2518}{81}
          - \frac{22 \pi^2}{27}
        \biggr) l_\mu
        - \frac{121}{27} l_\mu^2
      \biggr)
      \nonumber \\ & \mbox{}
      + C_A n_l T_F
      \biggl(
        \frac{19453}{486}
        - \frac{104 \pi^2}{81} 
        - 2 \zeta_3
        + \biggl(
          - \frac{1654}{81}
          + \frac{8\pi^2}{27}
        \biggr) l_\mu
        + \frac{88}{27} l_\mu^2
      \biggr)
      + C_F n_l T_F
      \biggl(
        \frac{11}{4}
        - \frac{4 \zeta_3}{3} 
        - \frac{2}{3} l_\mu
      \biggr)
      \nonumber \\ & \mbox{}
      + n_l^2 T_F^2
      \biggl(
        - \frac{1292}{243}
        + \frac{8\pi^2}{81}
        + \frac{256}{81} l_\mu
        - \frac{16}{27} l_\mu^2
      \biggr)
    \Biggr]
    + \frac{\mu^2}{\left({m^{\text{OS}}}\right)^2}
    \Biggl[
      C_A^2
      \biggl(
        - \frac{96295}{5184}
        + \frac{445 \pi^2}{432} 
        + \frac{57 \zeta_3}{16} 
        - \frac{\pi^4}{48}
        \nonumber \\ & \mbox{}
        + \biggl(
          \frac{2155}{216}
          - \frac{11 \pi^2}{36} 
        \biggr) l_\mu
        - \frac{121}{72} l_\mu^2
      \biggr)
      + C_A n_l T_F
      \biggl(
        \frac{13699}{1296}
        - \frac{23 \pi^2}{54}
        - \frac{3 \zeta_3}{4}
        + \biggl(
          - \frac{695}{108}
          + \frac{\pi^2}{9}
        \biggr) l_\mu
        + \frac{11}{9} l_\mu^2
      \biggr)
      \nonumber \\ & \mbox{}
      + C_F n_l T_F
      \biggl(
        \frac{29}{32}
        - \frac{\zeta_3}{2}
        - \frac{1}{4} l_\mu
      \biggr)
      + n_l^2 T_F^2
      \biggl(
        - \frac{209}{162}
        + \frac{\pi^2}{27}
        + \frac{26}{27} l_\mu
        - \frac{2}{9} l_\mu^2
      \biggr)
    \Biggr]
  \Biggr\}
  ~,
  \label{eq:7}
\end{align}
with $l_\mu = \ln \frac{2\mu}{\mu_s}$ ($\mu$ denotes the Wilsonian cutoff and $\mu_s$ the 
renormalization scale of the strong coupling constant)
and the $SU(N_c)$ color factors are given by $C_F=(N_c^2-1)/2N_c$, $C_A = N_c$ and $T_F=1/2$.
Note that this relation takes into account finite charm quark mass effects. These effects
are given by decoupling effects only, which we showed by explicit calculation.
%To obtain the relation between the kinetic and often used $\overline{\text{MS}}$ mass scheme 
%one can insert its known relation to the on-shell mass (see...)
%in Eq.~\eqref{eq:7}. 
The conversion between the kinetic mass and other mass schemes has been included in the 
public programs \texttt{RunDec} \cite{Herren:2017osy} and \texttt{REvolver} \cite{Hoang:2021fhn}.

\section{\boldmath The Semileptonic Decay Width to $\mathcal{O}(\alpha_s^3)$}

\begin{figure}
  \centering 
  \includegraphics[width=0.32\textwidth]{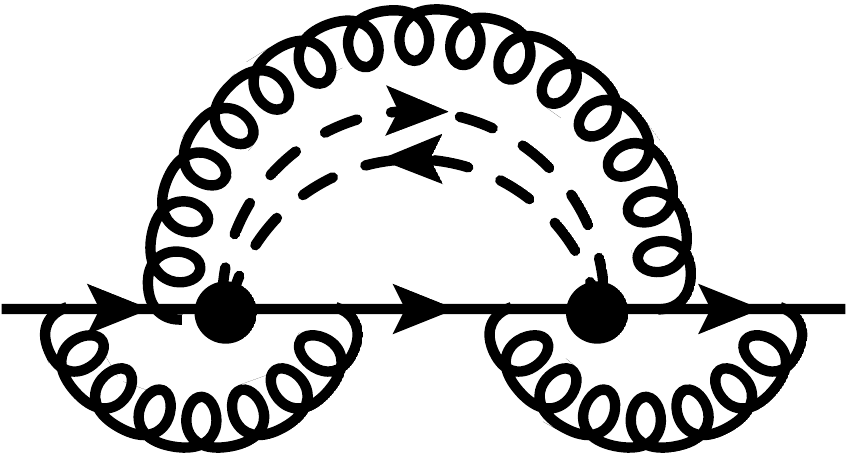}
  \includegraphics[width=0.32\textwidth]{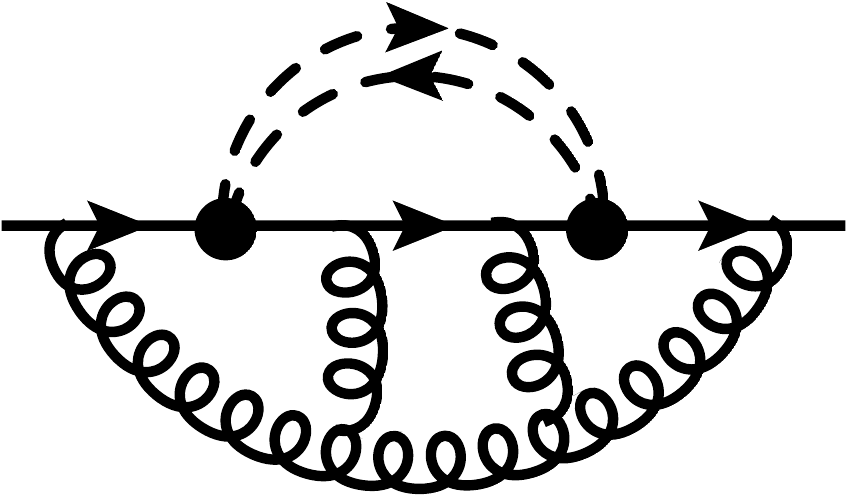}
  \includegraphics[width=0.32\textwidth]{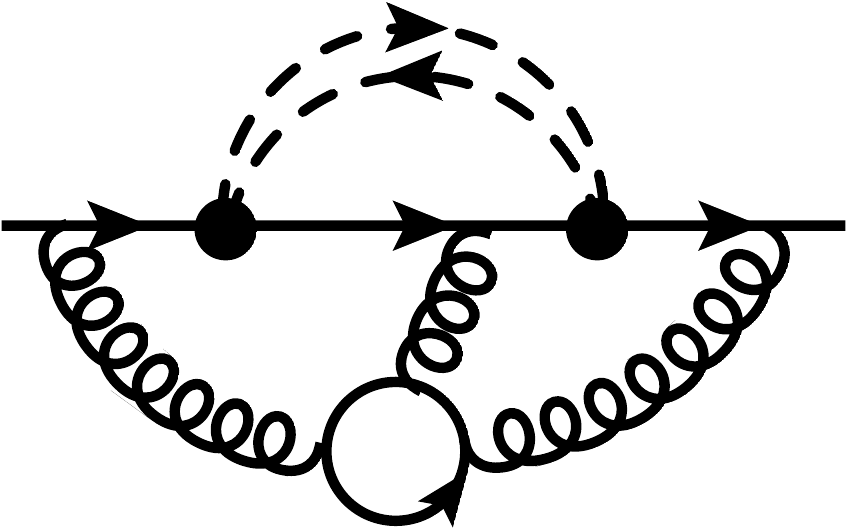}
  \caption{Sample Feynman diagrams for calculation of the semileptonic decay rate 
  at $\mathcal{O}(\alpha_s^3)$.
  Straight, curly and dashed lines represent quarks, gluons and leptons, respectively. 
  The weak interaction mediated by the $W$ boson is shown as a blob. 
  Taken from Ref.~\cite{Fael:2020tow}.}
  \label{fig:2}
\end{figure}

For the computation of the semileptonic decay width, we need to calculate the process
\begin{align}
  b(q) \to X_c(p_x) \ell(p_\ell) \overline{\nu}(p_\nu),
\end{align}
where $X_c$ is an inclusive state containing at least one charm quark and 
potentially other light quarks and gluons.
We can again use the optical theorem and consider the imaginary parts of 5-loop forward 
scattering diagrams (see Figure~\ref{fig:2}).

Since a calculation with complete analytical dependence on the charm and bottom mass 
seems out of reach, we consider the diagrams in an asymptotic expansion around 
\begin{align}
  \delta &= 1 - \frac{m_c}{m_b} 
  ~.
\end{align}
Although the expansion parameter is large for physical values of $m_c$ and $m_b$
($\delta \sim 0.7$), it has been shown in Ref.\cite{Dowling:2008mc} at $\mathcal{O}(\alpha_s^2)$
that this expansion converges well at the physical point and can even be extended
down to $m_c \to 0$ ($\delta \to 1$) with reasonable precision.
Furthermore, it turns out that in this limit the calculation simplifies:
\begin{itemize}
  \item For the asymptotic expansion in the limit $\delta \to 0$ we can use 
  expansion by regions. Here, the loop momenta can be either hard 
  $k_i \sim m_b$ or ultrasoft $k_i \sim \delta m_b$ again.
  \item The leptonic momenta have to be ultrasoft in order to generate an imaginary part.
  This reduces the number of regions to be considered.
  \item In the $\delta$-expansion one can completely factorize the leptonic system and 
  integrate it out without IBP reduction. We are therefore left with 3-loop integrals, 
  although we started from 5-loop diagrams. 
\end{itemize}
For the remaining 3-loop diagrams the scaling of the loop momenta can again either 
be hard or ultrasoft and the calculation can be performed in close analogy to the 
one of the kinetic mass relation discussed before.
However, since we are not only interested in the leading term of the expansion in $\delta$
but aim for 8 terms in the expansion, we encounter huge intermediate expressions of 
$\mathcal{O}(100 \, \text{GB})$ for individual diagrams and $\mathcal{O}(10^7)$ scalar 
integrals with positive and negative indices up to 12 which needed to be reduced to master
integrals.
\footnote{We thank A. Smirnov for providing a private version of \texttt{FIRE} which was essential for the reduction.}   
Furthermore, the wave function and mass renormalization constants at $\mathcal{O}(\alpha_s^3)$ 
allowing for two massive quarks, where only a few term in the expansion $m_c \to 0$ had been known 
analytically before (see Ref.~\cite{Bekavac:2007tk}), 
needed to be extended in order to renormalize the present 
calculation (see Ref.~\cite{Fael:2020bgs}).

Parametrizing the total decay rate as 
\begin{eqnarray}
  \Gamma &=& \Gamma_0 \left( X_0 
  + C_F \sum\limits_{i=1}^\infty \left( \frac{\alpha_s}{\pi} \right)^i X_i
  \right)  
\end{eqnarray}
we obtain the following contributions at $\mathcal{O}(\alpha_s^3)$ 
\begin{align}
  X_3 &=
  \delta^5 
  \biggl(
     \frac{266929}{810}
    -\frac{5248 a_4}{27}
    +\frac{2186 \pi ^2 \zeta _3}{45}
    -\frac{4094 \zeta_3}{45}
    -\frac{1544 \zeta _5}{9}
    -\frac{656 l_2^4}{81}
    +\frac{1336}{405} \pi ^2 l_2^2
    +\frac{44888\pi ^2 l_2}{135}
    \nonumber \\ & \mbox{}
    -\frac{9944 \pi ^4}{2025}
    -\frac{608201 \pi ^2}{2430}
  \biggr)
  + \mathcal{O}(\delta^6).
\end{align}
We used the notations $l_2 = \ln(2)$, $a_4 = \text{Li}_4(1/2)$ and $\zeta_i$ is 
Riemanns zeta function.
In the result above the color factors are specified to QCD and the renormalization scale 
$\mu_s = m_b$ has been chosen.
The full result with general color factors and expanded up to $\mathcal{O}(\delta^{12})$
can be found in the ancillary file to Ref.~\cite{Fael:2020tow}.
Recently the results of a subset of color factors has been confirmed up to $\mathcal{O}(\delta^9)$
in Ref.~\cite{Czakon:2021ybq}.

\section{Phenomenological Results}

Using the values $\alpha_s^{(5)}(M_Z) = 0.1179 $ \cite{Zyla:2020zbs},
$\overline{m}_c(3 \, \text{GeV}) = 993~\text{MeV}$ \cite{Chetyrkin:2017lif} and 
$\overline{m}_b(\overline{m}_b) = 4163~\text{MeV}$ \cite{Chetyrkin:2009fv},
we obtain 
\begin{align}
  m_b^{\rm kin}(\mu = 1 \, \text{GeV}) &= 
  (4163 + 259 + 78 + 26 \pm 13) \, \text{MeV}
  = ( 4526 \pm 13 ) \, \text{MeV}
  ~.
\end{align}
We estimate the error as half the $\mathcal{O}(\alpha_s^3)$ correction, which is 
also consistent with the residual scale uncertainty and known contributions 
in the large $\beta_0$ approximation at 4-loop.
The same approach at $\mathcal{O}(\alpha_s^2)$ leads to an uncertainty of 
$39 \, \text{MeV}$, the three-loop results therefore reduce the perturbative uncertainty by 
about a factor of two.

For the semileptonic decay rate in the on-shell scheme with $m_b^{\rm OS}=m_b=4.7 \, \text{GeV}$ and 
$m_c^{\rm OS}=m_c=1.3 \, \text{GeV}$ we obtain 
\begin{align}
  \Gamma(m_b,m_c) &= \Gamma_0 X_0 
  \left[
    1 
    - 1.72 \frac{\alpha_s^{(5)}}{\pi}
    - 13.09 \left( \frac{\alpha_s^{(5)}}{\pi} \right)^2
    -162.82  \left( \frac{\alpha_s^{(5)}}{\pi} \right)^3
  \right]
\end{align}
One observes the expected bad convergence of the perturbative series.
Using the kinetic scheme for the bottom quark and the $\overline{\text{MS}}$ scheme 
for the charm quark mass we obtain 
\begin{align}
  \Gamma(m_b^{\rm kin},\overline{m}_c(3 \, \text{GeV})) &= \Gamma_0 X_0 
  \left[
    1 
    - 1.67 \frac{\alpha_s^{(4)}}{\pi}
    - 7.25 \left( \frac{\alpha_s^{(4)}}{\pi} \right)^2
    -28.6  \left( \frac{\alpha_s^{(4)}}{\pi} \right)^3
  \right]
\end{align}
Similar improvements in der perturbative behavior are also observed using other 
threshold mass schemes for the bottom quark mass.
For a more detailed discussion see Ref.~\cite{Fael:2020tow}.

Both results have already been used to update the inclusive determination of $|V_{cb}|$
\cite{Bordone:2021oof}.
The inclusion of the presented $\mathcal{O}(\alpha_s^3)$ corrections resulted in a small 
shift of the central value but reduced the uncertainty due to the semileptonic width $\Gamma$
by a factor of two.

\acknowledgments \noindent
This research was supported by the Deutsche Forschungsgemeinschaft (DFG, German ResearchFoundation) 
under grant 396021762 — TRR 257 “Particle Physics Phenomenology after theHiggs Discovery”.

\bibliographystyle{JHEP}
\bibliography{bib}

\end{document}